\documentclass[12pt,preprint]{aastex}
\begin{document}
\newcommand{\up}[1]{\ifmmode^{\rm #1}\else$^{\rm #1}$\fi}
\newcommand{\zdot}{\makebox[0pt][l]{.}}
\newcommand{\upd}{\up{d}}
\newcommand{\uph}{\up{h}}
\newcommand{\upm}{\up{m}}
\newcommand{\ups}{\up{s}}
\newcommand{\arcd}{\ifmmode^{\circ}\else$^{\circ}$\fi}
\newcommand{\arcm}{\ifmmode{'}\else$'$\fi}
\newcommand{\arcs}{\ifmmode{''}\else$''$\fi}

\title{The Araucaria Project. The Distance to the Local Group Galaxy WLM
from Near-Infrared Photometry of Cepheid Variables
\footnote{Based on observations obtained with the ESO NTT for 
Programmes 077.D-0423 and 080.D-0047(B) and with the Magellan telescope
at the Las Campanas Observatory}
}
\author{Wolfgang Gieren}
\affil{Universidad de Concepci{\'o}n, Departamento de Fisica, Astronomy Group,
Casilla 160-C, Concepci{\'o}n, Chile}
\authoremail{wgieren@astro-udec.cl}
\author{Grzegorz Pietrzy{\'n}ski}
\affil{Universidad de Concepci{\'o}n, Departamento de Fisica, Astronomy
Group,
Casilla 160-C,
Concepci{\'o}n, Chile}
\affil{Warsaw University Observatory, Al. Ujazdowskie 4, 00-478, Warsaw,
Poland}
\authoremail{pietrzyn@astrouw.edu.pl}
\author{Olaf Szewczyk}
\affil{Universidad de Concepci{\'o}n, Departamento de Fisica, Astronomy
Group,
Casilla 160-C, Concepci{\'o}n, Chile}
\authoremail{szewczyk@astro-udec.cl}
\author{Igor Soszy{\'n}ski}
\affil{Warsaw University Observatory, Al. Ujazdowskie 4, 00-478, Warsaw,
Poland}
\authoremail{soszynsk@astrouw.edu.pl}
\author{Fabio Bresolin}
\affil{Institute for Astronomy, University of Hawaii at Manoa, 2680 Woodlawn 
Drive, 
Honolulu HI 96822, USA}
\authoremail{bresolin@ifa.hawaii.edu}
\author{Rolf-Peter Kudritzki}
\affil{Institute for Astronomy, University of Hawaii at Manoa, 2680 Woodlawn 
Drive,
Honolulu HI 96822, USA}
\authoremail{kud@ifa.hawaii.edu}
\author{Miguel A. Urbaneja}
\affil{Institute for Astronomy, University of Hawaii at Manoa, 2680 Woodlawn
Drive,
Honolulu HI 96822, USA}
\authoremail{urbaneja@ifa.hawaii.edu}
\author{Jesper Storm}
\affil{Astrophysikalisches Institut Potsdam, An der Sternwarte 16, D-14482
Potsdam, Germany}
\authoremail{jstorm@aip.de}
\author{Dante Minniti}
\affil{Departamento de Astronomia y Astrofisica, Pontificia Universidad Cat{\'o}lica
de Chile, Casilla 306, Santiago 22, Chile}
\authoremail{dante@astro.puc.cl}

\begin{abstract}
We have obtained deep images in the near-infrared J and K filters for several fields
in the Local Group galaxy WLM. We report intensity mean magnitudes for 31 Cepheids
located in these fields which we previously discovered in a wide-field optical imaging
survey of WLM. The data define tight period-luminosity relations in both near-infrared bands 
which we use to derive the total reddening of the Cepheids in WLM and the true distance modulus 
of the galaxy from a multiwavelength analysis of the reddened distance moduli
in the VIJK bands. From this, we
obtain the values E(B-V) = 0.082 $\pm$ 0.02, and $(m-M)_{0} = 24.924 \pm 0.042$
mag, with a systematic uncertainty in the distance of about $\pm$ 3\%. This Cepheid distance agrees extremely
well with the distance of WLM determined from the I-band TRGB method by ourselves
and others. Most of the reddening of the Cepheids in WLM (0.06 mag) is produced inside
the galaxy, demonstrating again the need for an accurate determination of the
total reddening and/or the use of infrared photometry to derive Cepheid
distances which are accurate to 3\% or better, even for small irregular galaxies
like WLM.  
\end{abstract}


\section{Introduction}

The effectiveness of using multiwavelength optical and near-infrared (NIR) observations
of Cepheid variables for distance determination of galaxies has been known for a long
time (McGonegal et al. 1982; Madore \& Freedman 1991). Only recently however the
technical problems with obtaining accurate and reliable NIR photometry for faint objects
in dense regions have been solved. Using NIR observations of Cepheids provides a number
of advantages for accurate distance work. First, the total and differential reddening
is significantly reduced in comparison to the optical bandpasses. Second, the Cepheid
period-luminosity (PL) relation becomes steeper toward longer wavelengths, and its intrinsic dispersion
becomes smaller (e.g. Fouqu{\'e} et al. 2007), both factors helping in deriving a more accurate distance. Third, 
metallicity effects on the PL relation in the near-IR are expected to be less important than at
optical wavelengths (Bono et al. 1999). Fourth, and very importantly from an observational
point of view, the amplitudes of variability are significantly smaller in the NIR
than at optical wavelengths, so even random single-epoch observations
approximate the mean magnitude reasonably well. If the period and optical light curve
of a Cepheid is accurately known, it is possible to derive its mean magnitude in the NIR bands
with an impressive 1-2\% accuracy from just one single-epoch observation (Soszy{\'n}ski
et al. 2005). 

Simultaneously studying NIR and optical PL relations of Cepheids provides another
important advantage. By combining the reddened distance moduli in the optical
and NIR it is possible to derive both the total reddening and the true distance modulus
with very good accuracy. This has been demonstrated in the previous papers of our ongoing
Araucaria Project which reported the distances to NGC 300 (Gieren et al. 2005b), IC 1613
(Pietrzy{\'n}ski et al. 2006), NGC 6822 (Gieren et al. 2006), NGC 3109 (Soszy{\'n}ski
et al. 2006) and NGC 55 (Gieren et al. 2008) 
derived by this method. For all these galaxies, we were able to determine
their distances {\it with respect to the LMC} with a 3-5\% accuracy from our technique. 
The possibility to 
measure the reddening produced {\it inside} these galaxies has been a fundamental step 
towards achieving this accuracy. We recall here that the principal purpose of the Araucaria
Project is to measure the distances to a sample of nearby galaxies with widely different
environmental properties with different stellar distance indicators, and compare the results
to determine the metallicity and age dependences of each standard candle.

In a previous paper (Pietrzy{\'n}ski et al. 2007; hereafter Paper I), 
we reported on an extensive wide-field imaging survey for Cepheids in the WLM irregular
galaxy carried out in the optical V and I bands in which we have discovered 60 variables, greatly extending 
the small sample of 15 Cepheids
which had been previously discovered in WLM by Sandage \& Carlson (1985). The WLM irregular galaxy
is an important target of the Araucaria Project because of its low metallicity
environment  which allows to study the behavior of stellar
distances indicators in the very low-metallicity regime. Indeed, the young stellar population of WLM
with a mean metallicity of -0.8 dex (Bresolin et al. 2006; Urbaneja  et al. 2008), is
slightly more metal-poor than that of the SMC.
 From the VI optical
photometry of the Cepheids we obtained in Paper I a true distance modulus of 25.144 mag for WLM.
In the course of the work for the present paper, we discovered that unfortunately
we had confused in Paper I the distance moduli for the V and Wesenheit bands; actually our
adopted best value for the WLM true distance modulus from the reddening-free Wesenheit index in Paper I 
is 25.014 mag (and not 25.144 mag). In order to improve on the distance value derived from optical
photometry of the Cepheids,
it was clearly in order to extend the Cepheid distance work to the near-infrared
domain and obtain a result which is unaffected by reddening. This
is the strategy we have been employing in all the Araucaria work
we have been conducting so far. In this paper, we therefore extend the
light curve coverage for 31 Cepheids in WLM to the NIR J and K bands. We then utilize the
multiband data available for these stars for an accurate determination of the total (average)
interstellar extinction to the Cepheids in WLM, and determine an improved Cepheid distance
to the galaxy from the multiwavelength method. 

The paper is organized as follows. In section 2 we describe the NIR observations, data reductions
and calibration of our photometry. In section 3 we derive the J- and K-band Cepheid PL relations
in WLM from our data and determine the true distance modulus to the galaxy from a multiwavelength
analysis. In section 4, we discuss our results, and in section 5 we summarize our conclusions.

\section{Observations, Data Reduction and Calibration}

The near-infrared data presented in this paper were collected during three observing runs in
2006 and 2007. We have been using two different telescopes and cameras for the data acqisition.
On September 22 and 23, 2006, and on November 24 and 25, 2007 we used the SOFI infrared camera
at the NTT at La Silla Observatory with a $4.9 \times 4.9$ arcmin field of view and a pixel scale of
0.288 arcsec/pixel, while on 22 and 23 November 2007 the PANIC infrared camera attached to the
Baade Magellan Telescope at Las Campanas Observatory was used. For the PANIC observations, the
field of view was $2.1 \times 2.1$ arcmin, and the pixel scale was 0.125 arcsec/pixel.

During the four nights at La Silla which were all photometric, we obtained deep $Js$ and $Ks$ 
observations of one field in WLM containing 23 Cepheid variables for which previous optical light curves
were presented in Paper I. With the PANIC camera we were able to observe three additional smaller fields partly
overlapping with the SOFI field, under photometric conditions.
The locations of all these fields are shown in Fig. 1 and their coordinates are given in Table 1.
Alltogether, the four observed fields in WLM contain 31 Cepheids.
In order to account for the rapid
variations of the sky brightness in the near-infrared bands we used a dithering technique,
as described in the previous papers of the Araucaria Project reporting infrared photometry
of Cepheid variables. Total integration times were 15 min in the $J$ and 60 min in the $K$ band.

All the reductions and calibrations were performed with the pipeline developed in the course
of the Araucaria Project and described in detail in earlier papers of this series. The subtraction of the
sky brightness was done in a two-step process which included the masking of stars with the
IRAF xdimsum package (Pietrzynski \& Gieren 2002). Next, each single image was flat-fielded and stacked 
into the final deep field image. PSF photometry and aperture corrections were performed as desribed in
Pietrzynski et al. (2002). The calibration of the photometry onto the standard system was based
on the observations of 22 standard stars from the UKIRT list (Hawarden et al. 2001). All of them
were observed along with the target fields in WLM at different airmasses and under photometric
conditions. The large number of standard stars observed on each of the six photometric nights
for this programme allowed us to obtain the absolute photometric zero points with an accuracy
close to 0.01 mag, in both filters. Deviations of the photometric zero points obtained independently for different
nights never exceeded 0.02 mag, for both SOFI and PANIC data. This is demonstrated in Fig. 2.
For an external check of our photometry, we compared our calibrated magnitudes with those
of the 2MASS catalog for the (small) sample of common stars, not finding any evidence for a significant
zero point difference in J and/or K (see Fig. 3).

We present the calibrated individual near-infrared magnitudes for all Cepheids  located in the observed fields 
in Table 2 which lists the stars' IDs, heliocentric Julian days of the observations (mid-integrations), and the
measurements in J and K with their standard deviations. Depending on the positions of the Cepheids
in the galaxy, the number of individual JK observations range from one to six. On average, we were able
to collect 3 JK observations per star. The WLM Cepheids observed in the near-infrared span a period
range from 54 down to 2.5 days.

\section{Near-Infrared Period-Luminosity Relations}

The intensity mean magnitudes of the Cepheids were derived by taking a straight average of the
individual random-phase fluxes measured of each variable, and converting the average fluxes back into
magnitudes. In principle a more accurate procedure to determine the mean magnitudes
in J and K would be the recipy described by Soszynski et al. (2005) which uses the V and I light
curves of the variables and the known phases of the near-infrared observations. However, in the
present case of WLM the epoch difference between the near-infrared data reported in this
paper, and the previous VI data reported in Paper I is so large, typically some 200 pulsation cycles,
 that with the limited
accuracy of the periods of the variables derived in Paper I the phasing of the near-IR data becomes very
uncertain. While this is unfortunate, the simple taking of a straight average of several random-phase 
magnitudes in a near-IR band of a Cepheid still does produce a rather accurate mean magnitude,
given the low light curve amplitudes of Cepheid variables at these wavelengths of typically
0.3 mag for stars with periods less than 10 days (e.g. Persson et al. 2004). For the one
long-period Cepheid in WLM, cep001 with a period of 54 days, the amplitude of the K-band
light curve is expected to be about 0.5 mag, but for this variable we have obtained six observations
at different phases which makes us expect that their mean value is very close to the
true mean magnitude of the variable, in the two bands we observed.

Table 3 gives the intensity mean J and K magnitudes of the individual Cepheids, with their
estimated uncertainties from the number and accuracy of the individual observations leading
to the adopted mean magnitude. We also provide the periods (adopted from Paper I). In Figures 4 and 5
we show the period-mean magnitude relations in the J and K bands as defined by the data
in Table 3. There is
one Cepheid, cep038, which is clearly over-luminous in both PL diagrams, by about 1.5 mag in J
and about 2 mag in K. We assume that the very bright magnitude of this variable is caused
by a nearby bright object which is not resolved in our images. Since the V and I magnitudes
of cep038 are normal for its period (see Paper I), the blend must be a very red object.
The ocurrence of at least one strongly blended Cepheid in WLM in a sample of about 30 stars is quite
expected from the result obtained by Bresolin et al. (2005) who studied the blending
of Cepheids in NGC 300, at about twice the distance of WLM,
 from a comparison of ground-based and HST/ACS photometry, finding three
strongly blended Cepheids in a sample of 16 Cepheids in this galaxy.  For the
following distance analysis, we exclude star cep038.

For the reasons discussed in Paper I, and in conformity with the approach adopted there,
we adopt a period cutoff of log P (days) = 0.5 for the distance analysis, retaining only the
Cepheids with longer periods in the sample. This ensures that
possible overtone pulsators are likely to be excluded in the sample adopted for
the distance determination, and it eliminates the variables with the lowest signal-to-noise
ratio in the photometry. The final sample consists of Cepheids 001-033 in Table 3 (24 stars). Note
that the strongly blended Cepheid cep038 is eliminated from the final sample also on the basis of the
adopted period cutoff.

>From the 24 Cepheids in the final sample, least-squares fits to a line yield slopes 
of -3.067 $\pm$ 0.204 in K, and -3.073 $\pm$ 0.233
in J, respectively. These slope values are shallower than, but within 1 $\sigma$ consistent with
the slopes for the Cepheid PL relations in the LMC, which are -3.261 in K, and -3.153 in J (Persson et al. 2004).
Following the procedure we have used in our previous papers, we adopt the LMC
slopes of Persson et al. (2004) in our fits. This yields the following PL relations for WLM in the J and K bands: \\

J = -3.153 log P + (22.795 $\pm$ 0.055)      $\sigma$ = 0.256 \\

K = -3.261 log P + (22.436 $\pm$ 0.048)      $\sigma$ = 0.223 \\

The dispersions are larger than the dispersions of the J and K band PL relations found in the LMC
by Persson et al. (2004), which are 0.12 mag in both bands. These values should closely resemble the
{\it intrinsic} dispersions of these relations, given that they are based on almost a hundred Cepheids
whose mean magnitudes were derived from full infrared light curves. The dispersions of the present
PL relations observed in WLM are larger mainly because they are based on a smaller number of stars,
and on mean magnitudes which are less precise than in the LMC work of Persson et al. due to the existence
of a few phase points only for each Cepheid.
In order to determine the relative distance moduli between WLM and the LMC, we need to convert the NICMOS (LCO)
photometric system used by Persson et al. (2004) to the UKIRT system utilized in this paper.
According to Hawarden et al. (2001),
there are just zero point offsets between the UKIRT and NICMOS systems (e.g. no color dependences)
in the J and K filters, which amount to 0.034 and 0.015 mag, respectively. Applying
these offsets, and assuming an LMC true distance modulus of 18.50 as in our previous
work in the Araucaria Project, we derive distance moduli for WLM of 24.993 $\pm$ 0.055 mag
in the J band, and 24.915 $\pm$ 0.045 mag in the K band.

As in our previous work, we will combine the distance moduli obtained from the near-infrared
photometry with the values we had previously derived in the optical VI bands in Paper I, to obtain
a very robust determination of both the true distance modulus of WLM, and the total (mean) reddening 
of the WLM Cepheids in our sample. When re-analyzing the results of Paper I in the course of this work, we detected that
by accident we had interchanged the distance modulus results reported for the V and Wesenheit bands (p. 601 of Paper I). The
correct values for the reddened distance moduli in V and I derived from our photometry in Paper I
are 25.156 $\pm$ 0.04 mag and 25.121 $\pm$ 0.03 mag, respectively. 
As in our previous papers in this series 
we adopt the extinction law of Schlegel et al. (1998) and fit a straight line to the relation
$(m-M)_{0} = (m-M)_{\lambda} - A_{\lambda} = (m-M)_{\lambda} - E(B-V) * R_{\lambda}$.
Using the corrected, reddened distance moduli in the V and I photometric bands as given above, together with
the values for the J and K bands obtained in this paper, we obtain the following values
for the reddening and the true distance modulus of WLM from the multiwavelength analysis: \\

$ E(B-V) = 0.082 \pm 0.020$

$(m-M)_{0} = 24.924 \pm 0.042$,

This corresponds to a distance of WLM of 0.97 $\pm$ 0.02 Mpc. 

In Table 4, we give the adopted values of $R_{\lambda}$ and the unreddened, true distance moduli
in each band which are obtained with the reddening value determined in our multi-wavelength
solution. The agreement between the unreddened distance moduli obtained in each band is very good.
In Fig. 6, we plot the apparent Cepheid distance moduli for WLM in VIJK as a function of $R_{\lambda}$, and the
best fitting straight line to the data. It is appreciated that the total reddening, and the true distance
modulus of WLM are indeed very well determined from this fit. By comparison with the foreground
reddening of 0.02 mag (Schlegel et al. 1998) it is seen that most of the total average reddening
of the WLM Cepheids is produced inside their host galaxy.

\section{Discussion}
As in the previous distance determinations of Local and Sculptor Group irregular and spiral galaxies 
from combined Cepheid near-infrared and optical photometry (see references cited in the Introduction)
obtained in the Araucaria Project, our combined near-infrared and optical data for a sample of
31 Cepheids in WLM have led to a very robust distance determination for this relatively distant 
Local Group galaxy. Again, as for the other galaxies we have studied so far, we find that the
slopes of the Cepheid PL relations in WLM in the near-infrared J and K bands are statistically in
 agreement with the one defined by the LMC Cepheids (Persson et al. 2004). Indeed, the free fits
 to the data discussed in section 3 of this paper yield slopes in both J and K which agree
 within 1 $\sigma$ with the adopted slope of Persson et al. (2004) from the LMC Cepheids. If we
 exclude the long-period Cepheid in the fits which carries a strong weight in the linear regression,
 the resulting values for the slopes of the PL relation in J and K become -2.868 $\pm$ 0.426,
 and -2.756 $\pm$ 0.368, respectively. Due to their large uncertainties, which is a consequence
 of the now very restricted period range of the remaining 23 Cepheid sample, these values still
 agree within 1.5 $\sigma$ with the slopes of the Cepheid PL relation as determined in the LMC
 by Persson et al. (2004). The observed slopes
 of the IR PL relations in WLM are therefore consistent  with the claim made by Gieren et al. (2005a)
and others that the slope of the Cepheid PL relation is metallicity-independent, at least
in near-infrared bands and for metallicities in the range
from -0.35 dex appropriate for the LMC Cepheids (Luck et al. 1998) down to -0.8 to -1.0 dex, which is the
metallicity  of the young stellar population in WLM (Urbaneja et al. 2008), IC 1613 (Bresolin et al. 2007)
and NGC 3109 (Evans et al. 2007) as determined from quantitative spectroscopy of blue supergiants.
These results from our project support our adopted procedure to use the extremely well-determined slopes
of the LMC Cepheid PL relations for the distance determination of our target galaxies. 

The distance of WLM derived in this paper is 0.09 mag shorter than the one derived in Paper I from
the Wesenheit index and agrees more closely with the distance for this galaxy derived from
the I-band magnitude of the tip of the red giant branch. Indeed, Lee et al. (1993) obtained
24.87 $\pm$ 0.08 mag for the distance of WLM from the TRGB method, while Minniti \& Zijlstra
measured 24.75 $\pm$ 0.1 from their data. More recent results from the same method are
24.85 $\pm$ 0.08 mag obtained by McConnachie et al. (2005), and
24.93 $\pm$ 0.04 mag, obtained by Rizzi et al. (2007).
Our own TRGB distance modulus of WLM reported in Paper I
is 24.91 $\pm$ 0.08. All these values are in excellent agreement with the WLM distance modulus
of 24.92 obtained in this paper from the Cepheid VIJK photometry. Our Cepheid distance modulus
is also in excellent agreement with the WLM distance modulus of 24.99 $\pm$ 0.10 mag
derived very recently by Urbaneja et al. (2008) from
a quantitative analysis of low resolution spectra of A and B supergiants via the flux-weighted
gravity-luminosity relation of Kudritzki et al. (2003, 2008). Other distance measurements reported
for WLM include the work of Valcheva et al. (2007) who obtained 24.84 $\pm$ 0.14 mag from IR photometry
of 4 Cepheids, Rejkuba et al. (2000) who measured 24.95 $\pm$ 0.13 mag from the V magnitude 
of the horizontal branch in WLM, and Dolphin (2000) who obtained 24.88 $\pm$ 0.09 mag from 
fitting the V, I color-magnitude diagram of the galaxy. All these distance measurements agree
within the uncertainties with the present Cepheid result. 

In the previous Cepheid
studies of our project in other Araucaria target galaxies the agreement between the distance
obtained from the VI Wesenheit index, and the full multiwavelength solution including
the near-infrared bands was usually better (in the order of 2 \%). Probably our result for WLM just demonstrates
that a truly reliable and robust distance determination from Cepheids definitively requires
the use of near-infrared photometry, and that the result from the (in principle) reddening-free
VI Wesenheit magnitude might yield a result which is still biased to some extent, particularly
when blending with unresolved stars in the optical images becomes a problem. Once we have
finished our VIJK distance analyses for all the target galaxies in our project, we will
analyze this problem more exhaustively.

The distance modulus of WLM derived in this paper is not affected by any significant amount
by the choice of the cutoff period adopted in the PL diagrams in Figs. 4 and 5. The
zero points in the WLM PL relations in both J and K change by less than 0.03 mag if we
use a longer period cutoff of log P (days) = 0.7 (leaving the 14 longest-period stars in the 
sample), or if we use the full sample of 30 Cepheids (excluding the one strongly
blended Cepheid cep038). Specifically, retaining the full Cepheid sample yields a distance
modulus of 24.902 $\pm$ 0.042 mag, whereas the sample of the 14 longest-period stars
yields 24.947 $\pm$ 0.042 mag. Both values agree within their statistical 1 $\sigma$ errors
with our adopted best distance modulus of 24.924 $\pm$ 0.042 mag. This probably demonstrates that 
our WLM Cepheid sample is large enough to fill the Cepheid instability strip in the HRD rather
homogeneously, and that there are no overtone Cepheids among the shortest period variables
in the sample, a conclusion which is supported by the asymmetric light curve shapes in V
of all these variables.

We estimate that the combined effect of the different sources which contribute to the
systematic uncertainty of our present multiwavelength Cepheid distance result for WLM
generate an uncertainty not exceeding $\pm$ 0.06 mag, or 3\%. These factors include
the accuracy of the photometric zero points, the effect of blending on the Cepheid
photometry, the effect of errors in the adopted reddening and a possible metallicity
dependence of the Cepheid PL relation. The typical impact of these factors on the
distance result has been discussed quite exhaustively in the previous papers of this
series. Here we only want to stress that the zero point of the near-infrared photometry
for WLM reported in this paper is probably even more accurate (0.01 mag) than for the
other, previously studied target galaxies of our project, as a consequence of having
data and independent photometric calibrations from 6 different photometric nights,
which has been an especially fortunate circumstance. The impact of blending of the
Cepheids on the distance was found to be less than 2\% in the case of NGC 300 (Bresolin et al. 2005)
which is at twice the distance of WLM, which allows us to conclude that its effect
in the present case of WLM is certainly not larger than 2\%, or 0.04 mag, and very likely smaller
than this. We adopt 0.03 mag for this source of systematic error.
Regarding the effect of reddening, our analysis and the results presented
in Fig. 6 and Table 4 convincingly demonstrate that the total reddening suffered by
the WLM Cepheids has been very accurately determined in this study, and any residual
effect of reddening on our distance result is clearly less than 2\%, or 0.04 mag. We want to stress
again that this elimination of reddening as a very serious source of systematic error
in Cepheid distances to galaxies based on optical photometry alone is probably
the single most important advantage offered by our combined optical-infrared 
procedure to measure Cepheid distances to late-type galaxies in our project. The influence
of metallicity effects on our distance result is harder to estimate, at the present time.
While we have presented evidence that the effect of metallicity differences on the slope of the
PL relation is probably negligible, its effect on the zero points of the PL relation
is not well established yet. We will investigate this question in detail in a forthcoming
paper in which we will compare the distance results for all our target galaxies from
the different stellar methods. However, previous work, like the one of Storm et al. (2004),
seem to indicate that the metallicity effect on the PL relation zero point is quite small.
We estimate its contribution to the systematic error of our WLM distance modulus to be
0.04 mag. Summing up the different contributions, we arrive at an estimated total $\pm$ 0.06 mag
systematic uncertainty of our present WLM distance modulus.

As in the previous papers of this series, our distance result for WLM is tied to an {\it assumed}
LMC distance modulus of 18.50 mag, which probably constitutes the single most important
source of systematic uncertainty (e.g. Benedict et al. 2002, Schaefer 2008). If
future work changes this adopted value of the LMC distance, we can easily adapt
the distances of the target galaxies of our project to the new value. The {\it relative}
distances between our different target galaxies will remain unaffected.

Finally, as remarked already before, our analyses of a number of galaxies with large differences
in the metallicities of their young stellar populations keep indicating us
so far that the slopes of the Cepheid PL relations in the various optical and
particularly in the near-infrared J and K bands are identical and thus not dependent
on metallicity. Subtle effects like the possible nonlinearity of the PL relation 
at 10 days as reported by Ngeow et al. (2008) do not seem to have any appreciable
effect on the distance measurement with this tool, particularly in the near-infrared domain.
The possible dependence of the zero points
of the PL relation in different bands on metallicity will be studied in detail
in a forthcoming paper of this series by comparison with the results from other
distance indicators.

\section{Conclusions}

The main conclusions of this paper can be summarized as follows:

1. We have obtained deep near-infrared photometry of a sizeable sample (31 stars)
of Cepheid variables in the Local Group galaxy WLM which were previously discovered
in a wide-field imaging survey conducted in the optical V and I bands.

2. From our calibrated infrared photometry in J and K we derive tight period-luminosity
relations for the Cepheids in WLM which are very well fitted by the slopes of the
LMC Cepheid PL relations in these bands of Persson et al. (2004), supporting the
conclusion that for a broad range of metallicities the Cepheid PL relation
in near-infrared bands is independent of metallicity.

3. Combining the reddened distance moduli in the J and K bands obtained in this paper
with those obtained in Paper I in the V and I bands we derive a total reddening
of the WLM Cepheids of E(B-V) = 0.082 $\pm$ 0.02, and a absorption corrected,
true distance modulus for WLM of 24.92 $\pm$ 0.04 mag (random error) $\pm$ 0.06 mag
(systematic error).

4. We report on an error in Paper I where we had confused the distance modulus
in the V band with the one in the Wesenheit band. For our present distance determination,
we have obviously used the correct value of the reddened distance modulus in V as obtained
from the data in Paper I. The true distance modulus from the Wesenheit magnitude
obtained in that paper must read 25.014 $\pm$ 0.036 mag, rather than the erroneosly
given value of 25.144 mag. 

5. The WLM distance derived from the multiwavelength VIJK Cepheid analysis in this
paper is in excellent agreement with the various determinations of the WLM distance
from the I-band tip of the red giant branch method and from other techniques
which have been reported in the literature.

6. With the completion of the Cepheid multiwavelength distance determination for WLM
there are now six late-type galaxies in the Araucaria Project for which such
distance determinations have been carried out (WLM, IC 1613, NGC 3109 and NGC 6822 in
the Local Group and NGC 300 and NGC 55 in the Sculptor Group; see references in the
Introduction). Corresponding work for two more Sculptor Group galaxies, NGC 247
and NGC 7793, is underway. A comparison of these distances with those derived for
the same galaxies from the other techniques we are using in our project
will be conducted in due time and is expected to lead to an improved determination
of the metallicity dependence of the different techniques. This will finally yield a 
set of very accurate distances to a number of nearby galaxies in the 0.05-4 Mpc
range which can be used to calibrate more accurately other techniques which
reach out to distances large enough for an improved determination of the
Hubble constant.

\acknowledgments
WG, GP and DM gratefully acknowledge financial support for this
work from the Chilean Center for Astrophysics FONDAP 15010003, and from
the BASAL Centro de Astrofisica y Tecnologias Afines (CATA). 
Support from the Polish grant N203 002 31/046 and the FOCUS 
subsidy of the Fundation for Polish Science (FNP)
is also acknowledged. It is a great pleasure
to thank the support astronomers at ESO-La Silla and at Las Campanas Observatory
for their expert help in the observations. We also thank the ESO OPC and CNTAC
for allotting generous amounts of observing time to this project.
Helpful comments of a referee on a previous version of this paper
are appreciated.

\begin{figure}[p] 
\vspace*{18cm}
\includegraphics{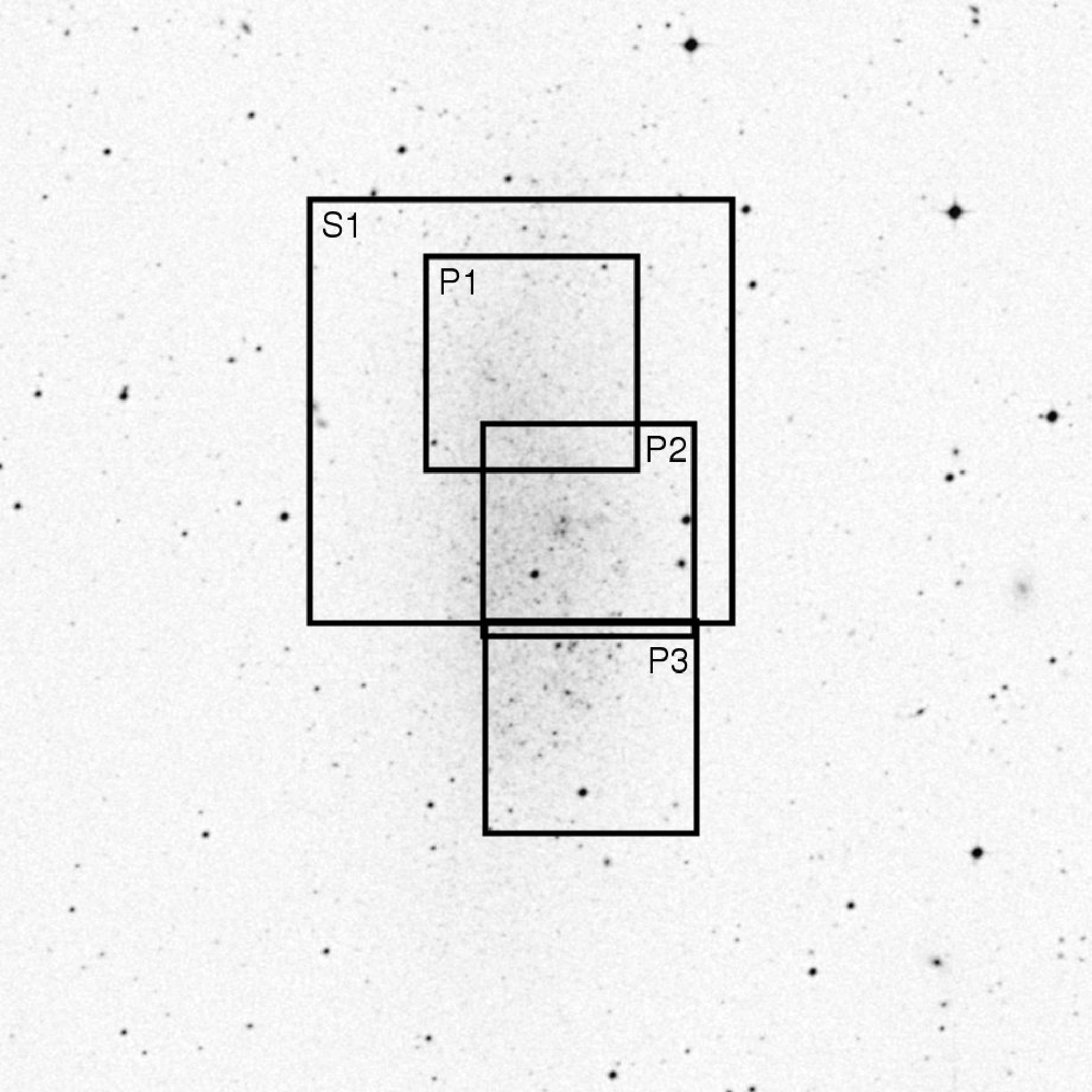} 
\caption{The location of the observed fields in WLM on the DSS
blue plate. We observed one NTT/SOFI field, and three Magellan/PANIC fields (see text).
The fields contain 31 Cepheid
variables. The size of the SOFI field (S1) is $4.9 \times 4.9$ arcmin.}
\end{figure}

\begin{figure}[p]
\vspace*{18cm}
\includegraphics{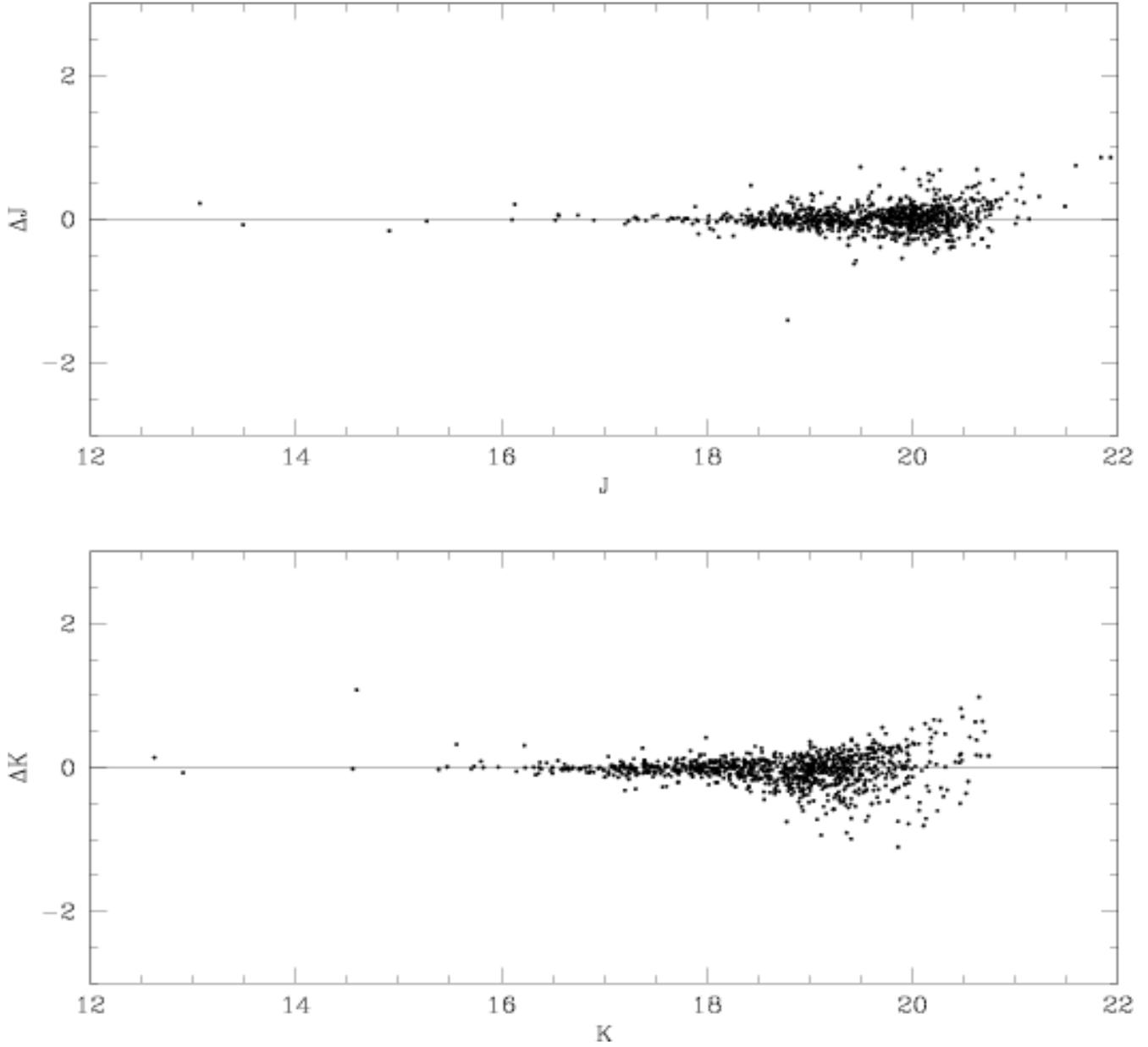}
\caption{An example of the comparison of the photometric zero points 
in the K and J bands of WLM obtained with the NTT 
and SOFI  on two different nights and calibrated
completely independently. The  difference in zero points for the data 
obtained for other nights with SOFI and PANIC are very similar and 
are always smaller than 0.02 mag, which reflects the very good quality of 
our calibrations.
}
\end{figure}

\begin{figure}[p]
\vspace*{18cm}
\includegraphics{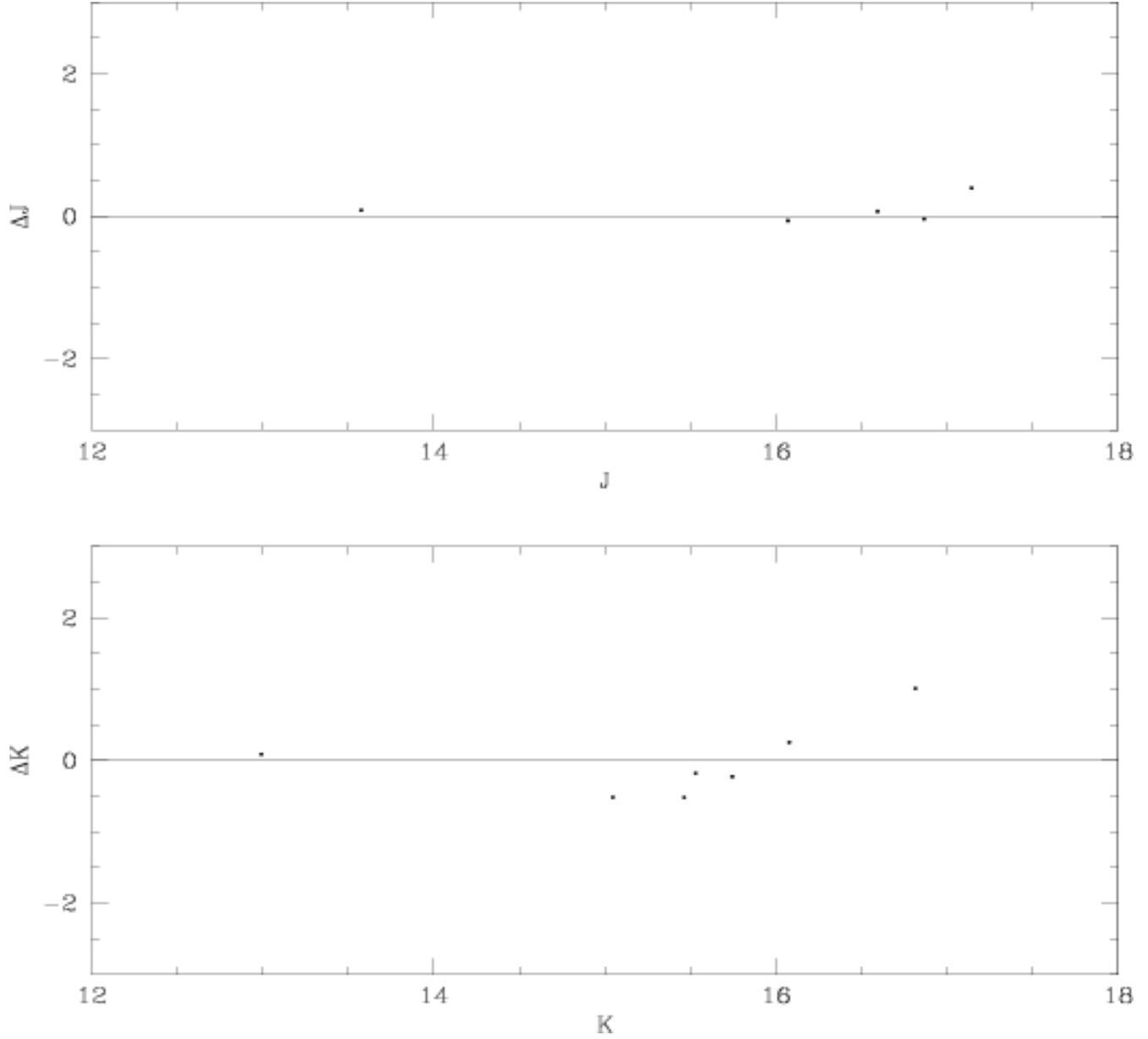}
\caption{Comparison of our photometry with the 2MASS data. In spite 
of the large scatter towards the fainter magnitudes caused by the 
low accuracy of the 2MASS photometry of faint stars, no clear evidence 
for significant offsets in the zero points is present.
}
\end{figure}

\begin{figure}[htb]
\vspace*{15cm}
\includegraphics{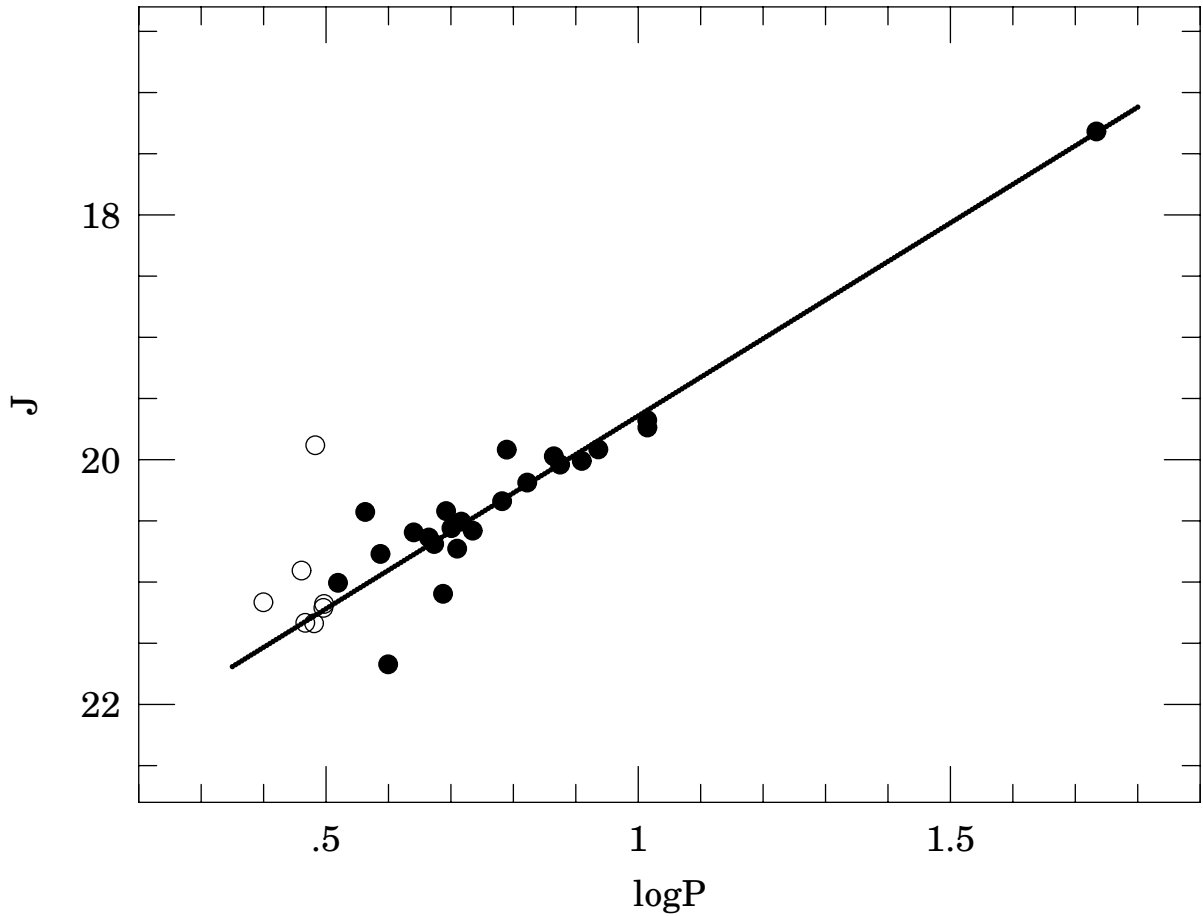}
\caption{Cepheid period-luminosity relations for WLM in the J  band. The 24 variables
with periods larger than 3.16 days (filled circles) have been adopted for the
distance determination. The slope of the fit to a line has been adopted from
the work of Persson et al. on the LMC Cepheids, and fits the WLM Cepheid data
very well.}
\end{figure}

\begin{figure}[htb]
\vspace*{15cm}
\includegraphics{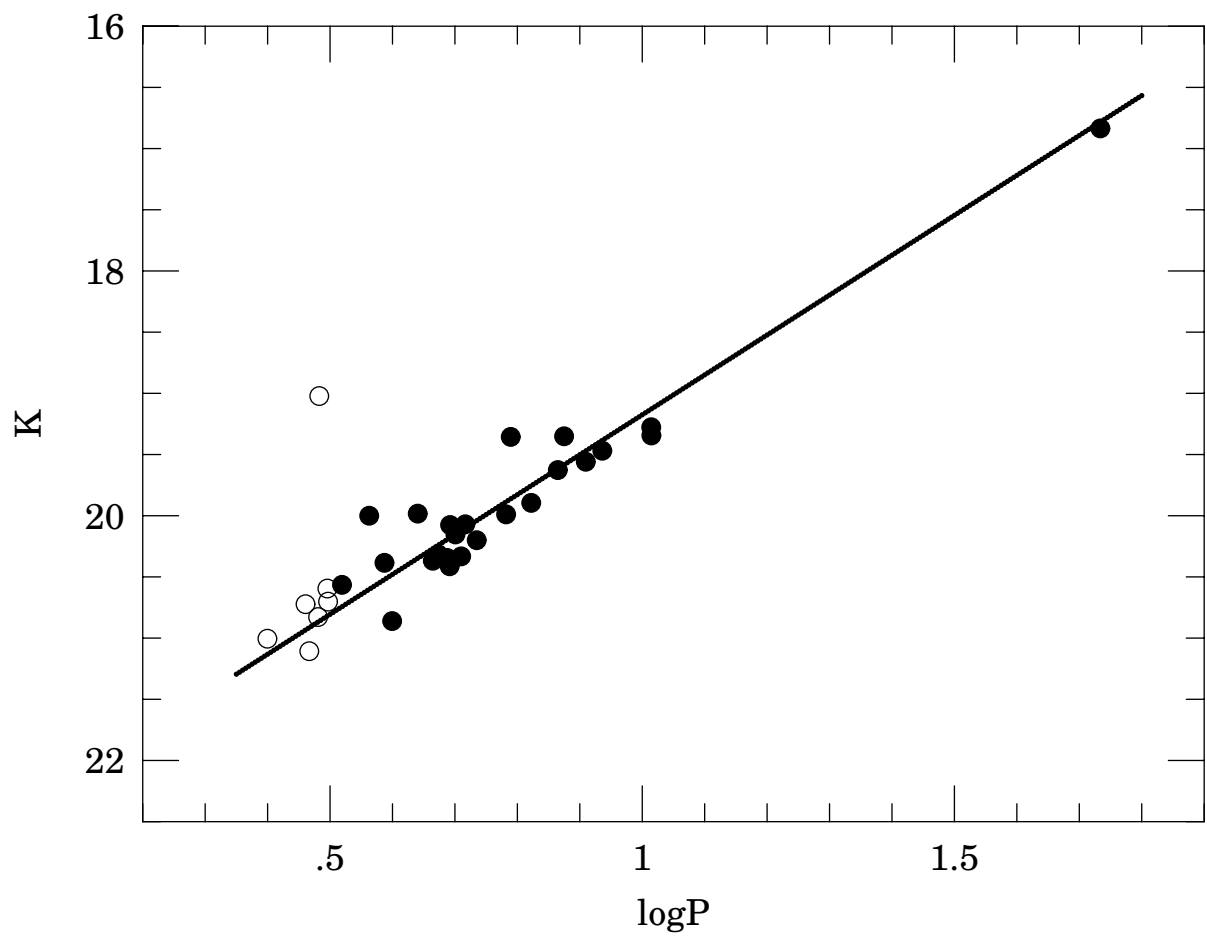}
\caption{Same as Fig. 4, for the K band.}
\end{figure}

\begin{figure}[p]
\includegraphics{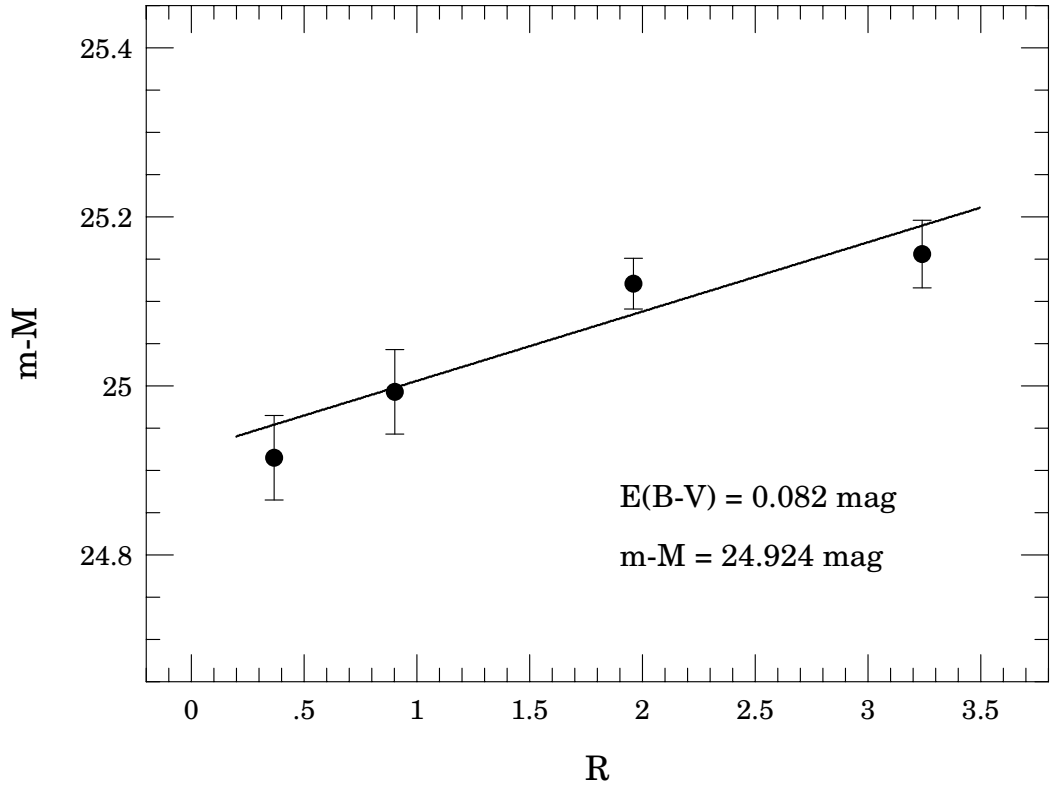}
\vspace{10cm}
\caption{Apparent distance moduli to WLM as derived in different photometric bands,
plotted against the ratio of total to selective extinction as adopted from
the Schlegel et al. reddening law. The intersection and
slope of the best-fitting line give the true distance modulus of WLM, and the
total mean reddening of its Cepheids, respectively. 
}
\end{figure}

\clearpage
\begin{deluxetable}{ccccc}
\tablecaption{}
\tablehead{\colhead{Field ID} & \colhead{Field name} & \colhead{RA 2000} 
& \colhead{DEC 2000} & \colhead{Instrument} }
\startdata
S1 & WLM-F1 & 00:01:57.67 & -15:25:29.5 & SOFI \\
P1 & WLM-F1 & 00:01:58.26 & -15:24:40.2 & PANIC \\
P2 & WLM-F2 & 00:01:54.98 & -15:27:01.8 & PANIC \\
P3 & WLM-F3 & 00:01:54.87 & -15:29:48.9 & PANIC \\
\enddata
\end{deluxetable}

\begin{deluxetable}{ccccccccc}
\rotate
\tabletypesize{\footnotesize}
\tablecaption{Journal of the Individual J and K band Observations of WLM}
\tablehead{\colhead{Star ID} & \colhead{Field ID} & \colhead{Period} & \colhead{J MJD} & \colhead{J} & \colhead{$\sigma$} & \colhead{K MJD} & \colhead{K} & \colhead{$\sigma$} \\ 
\colhead{} & \colhead{} & \colhead{(days)} & \colhead{-2400000} & \colhead{(mag)} & \colhead{(mag)} & \colhead{-2400000} & \colhead{(mag)} & \colhead{(mag)} } 
\startdata
cep001 & S1 & 54.17118 & 54001.08144853 & 17.309 & 0.024 & 54001.09783376 & 16.955 & 0.018 \\
cep001 & S1 & 54.17118 & 54002.10563506 & 17.327 & 0.019 & 54002.11759493 & 16.802 & 0.017 \\
cep001 & P1 & 54.17118 & 54427.06900000 & 17.341 & 0.004 & 54427.08676000 & 16.781 & 0.007 \\
cep001 & P1 & 54.17118 & 54428.14376000 & 17.326 & 0.004 & 54428.15841000 & 16.854 & 0.006 \\
cep001 & S1 & 54.17118 & 54429.01840384 & 17.313 & 0.014 & 54429.03095727 & 16.813 & 0.027 \\
cep001 & S1 & 54.17118 & 54430.01460106 & 17.292 & 0.013 & 54430.02712407 & 16.807 & 0.023 \\
cep002 & P3 & 10.34249 & 54428.06006000 & 19.737 & 0.020 & 54428.07505000 & 19.344 & 0.034 \\
cep003 & P2 & 10.33645 & 54427.15887000 & 19.677 & 0.025 & 54427.17380000 & 19.277 & 0.062 \\
cep005 & P3 & 8.63110 & 54428.06006000 & 19.916 & 0.020 & 54428.07505000 & 19.469 & 0.039 \\
cep007 & S1 & 8.12051 & 54001.08144853 & 19.927 & 0.083 & 54001.09783376 & 19.463 & 0.077 \\
cep007 & S1 & 8.12051 & 54002.10563506 & 19.984 & 0.063 & 54002.11759493 & 19.561 & 0.100 \\
cep007 & P1 & 8.12051 & 54427.06900000 & 20.182 & 0.019 & 54427.08676000 & 19.814 & 0.048 \\
cep007 & P1 & 8.12051 & 54428.14376000 & 20.126 & 0.022 & 54428.15841000 & 19.620 & 0.054 \\
cep007 & S1 & 8.12051 & 54429.01840384 & 19.840 & 0.078 & 54429.03095727 & 19.511 & 0.192 \\
cep007 & S1 & 8.12051 & 54430.01460106 & 20.001 & 0.104 & 54430.02712407 & 19.383 & 0.150 \\
cep008 & P2 & 7.49672 & 54427.15887000 & 20.040 & 0.029 & 54427.17380000 & 19.351 & 0.048 \\
cep010 & S1 & 7.32485 & 54001.08144853 & 20.167 & 0.092 & 54001.09783376 & 19.858 & 0.088 \\
cep010 & S1 & 7.32485 & 54002.10563506 & 19.868 & 0.050 & 54002.11759493 & 19.472 & 0.107 \\
cep010 & P1 & 7.32485 & 54427.06900000 & 19.955 & 0.024 & 54427.08676000 & 19.654 & 0.054 \\
cep010 & P1 & 7.32485 & 54428.14376000 & 19.899 & 0.030 & 54428.15841000 & 19.525 & 0.056 \\
cep011 & P3 & 6.64055 & 54428.06006000 & 20.187 & 0.018 & 54428.07505000 & 19.895 & 0.042 \\
cep012 & S1 & 6.15754 & 54001.08144853 & 19.841 & 0.083 & 54001.09783376 & 19.263 & 0.065 \\
cep012 & S1 & 6.15754 & 54002.10563506 & 19.930 & 0.071 & 54002.11759493 & 19.388 & 0.087 \\
cep012 & P1 & 6.15754 & 54427.06900000 & 99.999 & 9.999 & 54427.08676000 & 19.438 & 0.051 \\
cep012 & P1 & 6.15754 & 54428.14376000 & 20.131 & 0.040 & 54428.15841000 & 19.549 & 0.054 \\
cep012 & S1 & 6.15754 & 54429.01840384 & 19.770 & 0.075 & 54429.03095727 & 19.141 & 0.147 \\
cep013 & P3 & 6.05309 & 54428.06006000 & 20.339 & 0.027 & 54428.07505000 & 19.989 & 0.062 \\
cep015 & S1 & 5.43153 & 54001.08144853 & 20.626 & 0.124 & 54001.09783376 & 20.284 & 0.138 \\
cep015 & S1 & 5.43153 & 54002.10563506 & 20.638 & 0.091 & 54002.11759493 & 20.322 & 0.190 \\
cep015 & P1 & 5.43153 & 54427.06900000 & 20.574 & 0.028 & 54427.08676000 & 20.320 & 0.093 \\
cep015 & P1 & 5.43153 & 54428.14376000 & 20.677 & 0.036 & 54428.15841000 & 20.280 & 0.087 \\
cep015 & S1 & 5.43153 & 54430.01460106 & 20.386 & 0.131 & 54430.02712407 & 19.793 & 0.209 \\
cep016 & S1 & 5.20796 & 54001.08144853 & 20.664 & 0.120 & 54001.09783376 & 20.099 & 0.102 \\
cep016 & S1 & 5.20796 & 54002.10563506 & 20.582 & 0.071 & 54002.11759493 & 20.016 & 0.131 \\
cep016 & P1 & 5.20796 & 54427.06900000 & 20.385 & 0.022 & 54427.08676000 & 20.090 & 0.078 \\
cep016 & P1 & 5.20796 & 54428.14376000 & 20.493 & 0.035 & 54428.15841000 & 19.980 & 0.071 \\
cep016 & S1 & 5.20796 & 54430.01460106 & 20.404 & 0.134 & 54430.02712407 & 20.163 & 0.274 \\
cep017 & P3 & 5.12851 & 54428.06006000 & 20.727 & 0.025 & 54428.07505000 & 20.332 & 0.060 \\
cep018 & S1 & 5.02134 & 54001.08144853 & 20.555 & 0.118 & 54001.09783376 & 20.402 & 0.140 \\
cep018 & S1 & 5.02134 & 54002.10563506 & 20.567 & 0.080 & 54002.11759493 & 20.328 & 0.173 \\
cep018 & S1 & 5.02134 & 54429.01840384 & 20.555 & 0.149 & 54429.03095727 & 19.730 & 0.226 \\
cep019 & S1 & 4.92341 & 54001.08144853 & 20.338 & 0.107 & 54001.09783376 & 20.217 & 0.138 \\
cep019 & S1 & 4.92341 & 54002.10563506 & 20.494 & 0.086 & 54002.11759493 & 19.990 & 0.141 \\
cep019 & P1 & 4.92341 & 54427.06900000 & 20.379 & 0.031 & 54427.08676000 & 19.964 & 0.070 \\
cep019 & P1 & 4.92341 & 54428.14376000 & 20.626 & 0.037 & 54428.15841000 & 20.379 & 0.098 \\
cep019 & S1 & 4.92341 & 54429.01840384 & 20.265 & 0.113 & 54429.03095727 & 19.828 & 0.260 \\
cep020 & S1 & 4.91559 & 54429.01840384 & 99.999 & 9.999 & 54429.03095727 & 20.413 & 0.409 \\
cep021 & S1 & 4.86831 & 54002.10563506 & 21.096 & 0.115 & 54002.11759493 & 20.345 & 0.186 \\
cep022 & S1 & 4.71140 & 54001.08144853 & 20.726 & 0.136 & 54001.09783376 & 20.232 & 0.141 \\
cep022 & S1 & 4.71140 & 54002.10563506 & 20.735 & 0.097 & 54002.11759493 & 20.234 & 0.160 \\
cep022 & P1 & 4.71140 & 54427.06900000 & 20.782 & 0.040 & 54427.08676000 & 20.301 & 0.079 \\
cep022 & P1 & 4.71140 & 54428.14376000 & 20.511 & 0.032 & 54428.15841000 & 20.492 & 0.098 \\
cep023 & S1 & 4.61967 & 54001.08144853 & 20.597 & 0.122 & 54001.09783376 & 20.607 & 0.171 \\
cep023 & S1 & 4.61967 & 54002.10563506 & 20.669 & 0.093 & 54002.11759493 & 20.222 & 0.177 \\
cep023 & P1 & 4.61967 & 54427.06900000 & 20.630 & 0.026 & 54427.08676000 & 20.316 & 0.085 \\
cep023 & P1 & 4.61967 & 54428.14376000 & 20.646 & 0.039 & 54428.15841000 & 20.328 & 0.086 \\
cep024 & P2 & 4.36958 & 54427.15887000 & 20.594 & 0.048 & 54427.17380000 & 20.307 & 0.148 \\
cep024 & S1 & 4.36958 & 54430.01460106 & 99.999 & 9.999 & 54430.02712407 & 19.658 & 0.174 \\
cep026 & S1 & 3.97606 & 54001.08144853 & 21.726 & 0.299 & 54001.09783376 & 21.334 & 0.333 \\
cep026 & S1 & 3.97606 & 54002.10563506 & 21.617 & 0.180 & 54002.11759493 & 21.026 & 0.307 \\
cep026 & S1 & 3.97606 & 54429.01840384 & 99.999 & 9.999 & 54429.03095727 & 20.454 & 0.468 \\
cep026 & S1 & 3.97606 & 54430.01460106 & 99.999 & 9.999 & 54430.02712407 & 20.631 & 0.387 \\
cep027 & P2 & 3.86503 & 54427.15887000 & 20.770 & 0.048 & 54427.17380000 & 20.654 & 0.178 \\
cep027 & S1 & 3.86503 & 54429.01840384 & 99.999 & 9.999 & 54429.03095727 & 20.116 & 0.335 \\
cep031 & S1 & 3.65377 & 54001.08144853 & 20.290 & 0.092 & 54001.09783376 & 20.067 & 0.098 \\
cep031 & S1 & 3.65377 & 54002.10563506 & 20.487 & 0.064 & 54002.11759493 & 20.049 & 0.126 \\
cep031 & P1 & 3.65377 & 54427.06900000 & 20.401 & 0.029 & 54427.08676000 & 19.914 & 0.072 \\
cep031 & P1 & 3.65377 & 54428.14376000 & 20.533 & 0.045 & 54428.15841000 & 20.201 & 0.102 \\
cep031 & S1 & 3.65377 & 54429.01840384 & 20.684 & 0.154 & 54429.03095727 & 20.014 & 0.290 \\
cep031 & S1 & 3.65377 & 54430.01460106 & 20.169 & 0.106 & 54430.02712407 & 19.756 & 0.181 \\
cep033 & S1 & 3.30475 & 54001.08144853 & 99.999 & 9.999 & 54001.09783376 & 20.533 & 0.155 \\
cep033 & P1 & 3.30475 & 54428.14376000 & 21.006 & 0.074 & 54428.15841000 & 20.300 & 0.112 \\
cep033 & S1 & 3.30475 & 54430.01460106 & 99.999 & 9.999 & 54430.02712407 & 20.861 & 0.512 \\
cep036 & S1 & 3.13908 & 54001.08144853 & 21.178 & 0.182 & 54001.09783376 & 20.372 & 0.136 \\
cep036 & P1 & 3.13908 & 54427.06900000 & 21.180 & 0.051 & 54427.08676000 & 21.031 & 0.166 \\
cep037 & S1 & 3.13102 & 54001.08144853 & 21.301 & 0.205 & 54001.09783376 & 20.643 & 0.169 \\
cep037 & P1 & 3.13102 & 54427.06900000 & 21.067 & 0.040 & 54427.08676000 & 20.774 & 0.123 \\
cep037 & P1 & 3.13102 & 54428.14376000 & 21.264 & 0.058 & 54428.15841000 & 20.769 & 0.136 \\
cep037 & S1 & 3.13102 & 54429.01840384 & 99.999 & 9.999 & 54429.03095727 & 20.195 & 0.357 \\
cep038 & P1 & 3.03891 & 54427.06900000 & 19.839 & 0.018 & 54427.08676000 & 18.954 & 0.029 \\
cep038 & P1 & 3.03891 & 54428.14376000 & 19.834 & 0.019 & 54428.15841000 & 19.022 & 0.028 \\
cep038 & S1 & 3.03891 & 54429.01840384 & 19.957 & 0.084 & 54429.03095727 & 19.021 & 0.132 \\
cep038 & S1 & 3.03891 & 54430.01460106 & 19.899 & 0.093 & 54430.02712407 & 19.091 & 0.118 \\
cep039 & S1 & 3.02576 & 54001.08144853 & 21.155 & 0.184 & 54001.09783376 & 21.439 & 0.374 \\
cep039 & P1 & 3.02576 & 54427.06900000 & 21.061 & 0.044 & 54427.08676000 & 20.562 & 0.164 \\
cep039 & S1 & 3.02576 & 54429.01840384 & 21.796 & 0.431 & 54429.03095727 & 20.483 & 0.428 \\
cep041 & P3 & 2.92899 & 54428.06006000 & 21.333 & 0.072 & 54428.07505000 & 21.107 & 0.154 \\
cep043 & P3 & 2.88896 & 54428.06006000 & 20.906 & 0.051 & 54428.07505000 & 20.723 & 0.130 \\
cep049 & S1 & 2.51004 & 54001.08144853 & 21.440 & 0.235 & 54001.09783376 & 21.382 & 0.377 \\
cep049 & P1 & 2.51004 & 54427.06900000 & 20.947 & 0.037 & 54427.08676000 & 20.796 & 0.133 \\
cep049 & P1 & 2.51004 & 54428.14376000 & 21.108 & 0.052 & 54428.15841000 & 20.838 & 0.127 \\
\enddata
\end{deluxetable}

\begin{deluxetable}{cccccc}
\tablecaption{Intensity mean J and K magnitudes for 30 Cepheid variables in WLM}
\tablehead{\colhead{Star ID} & \colhead{Period} & \colhead{J} & \colhead{$\sigma$} & \colhead{K} & \colhead{$\sigma$} \\ 
\colhead{} & \colhead{(days)} & \colhead{(mag)} & \colhead{(mag)} & \colhead{(mag)} & \colhead{(mag)} } 
\startdata
cep001 & 54.17118 & 17.318 & 0.015 & 16.835 & 0.018 \\
cep002 & 10.34249 & 19.737 & 0.020 & 19.344 & 0.034 \\
cep003 & 10.33645 & 19.677 & 0.025 & 19.277 & 0.062 \\
cep005 & 8.63110 & 19.916 & 0.020 & 19.469 & 0.039 \\
cep007 & 8.12051 & 20.010 & 0.069 & 19.559 & 0.116 \\
cep008 & 7.49672 & 20.040 & 0.029 & 19.351 & 0.048 \\
cep010 & 7.32485 & 19.972 & 0.056 & 19.627 & 0.079 \\
cep011 & 6.64055 & 20.187 & 0.018 & 19.895 & 0.042 \\
cep012 & 6.15754 & 19.918 & 0.062 & 19.356 & 0.088 \\
cep013 & 6.05309 & 20.339 & 0.027 & 19.989 & 0.062 \\
cep015 & 5.43153 & 20.580 & 0.093 & 20.200 & 0.152 \\
cep016 & 5.20796 & 20.506 & 0.088 & 20.070 & 0.151 \\
cep017 & 5.12851 & 20.727 & 0.025 & 20.332 & 0.060 \\
cep018 & 5.02134 & 20.559 & 0.119 & 20.153 & 0.183 \\
cep019 & 4.92341 & 20.420 & 0.082 & 20.076 & 0.156 \\
cep020 & 4.91559 & 99.999 & 9.999 & 20.413 & 0.409 \\
cep021 & 4.86831 & 21.096 & 0.115 & 20.345 & 0.186 \\
cep022 & 4.71140 & 20.689 & 0.087 & 20.315 & 0.124 \\
cep023 & 4.61967 & 20.636 & 0.080 & 20.368 & 0.137 \\
cep024 & 4.36958 & 20.594 & 0.034 & 19.983 & 0.162 \\
cep026 & 3.97606 & 21.672 & 0.175 & 20.861 & 0.379 \\
cep027 & 3.86503 & 20.770 & 0.034 & 20.385 & 0.268 \\
cep031 & 3.65377 & 20.427 & 0.092 & 20.000 & 0.162 \\
cep033 & 3.30475 & 21.006 & 0.043 & 20.565 & 0.316 \\
cep036 & 3.13908 & 21.179 & 0.134 & 20.702 & 0.152 \\
cep037 & 3.13102 & 21.211 & 0.108 & 20.595 & 0.218 \\
cep038 & 3.03891 & 19.882 & 0.064 & 19.022 & 0.091 \\
cep039 & 3.02576 & 21.337 & 0.272 & 20.828 & 0.342 \\
cep041 & 2.92899 & 21.333 & 0.072 & 21.107 & 0.154 \\
cep043 & 2.88896 & 20.906 & 0.051 & 20.723 & 0.130 \\
cep049 & 2.51004 & 21.165 & 0.141 & 21.005 & 0.242 \\
\enddata
\end{deluxetable}

\begin{deluxetable}{cccccc}
\tablewidth{0pc}
\tablecaption{Reddened and Absorption-Corrected Distance Moduli for
WLM in Optical and Near-Infrared Bands}
\tablehead{ \colhead{Band} & $V$ & $I$ & $J$ & $K$ & $E(B-V)$ }
\startdata
 $m-M$                &   25.156 &  25.121 &  24.993 &  24.915 &   --  \nl
 ${\rm R}_{\lambda}$  &   3.24   &  1.96   &  0.902  &  0.367  &   --  \nl
$(m-M)_{0}$           &  24.896 &  24.964 &  24.920  &  24.885 &  0.08 \nl
\enddata
\end{deluxetable}

\end{document}